\documentclass[12pt]{article}
\usepackage{citesort}
\usepackage{epsfig}
\usepackage{amsmath}
\usepackage{graphicx}

\addtocounter{footnote}{-1}

\date{Oct.25.01}

\begin{document}           
\title{ \vskip -1.0in
\Large \bf{ Protein folding mediated by solvation:  \\
water expelling  \\ and formation of the hydrophobic core 
occurs after the structural collapse}}
\vskip 3.5cm
\author{Margaret S. Cheung$^*$, Angel E. Garc\'{\i}a$^{\dagger}$,
and Jos\'e N. 
Onuchic$^*$ \\[16pt]
Department of Physics, University of California at 
San Diego,\\
La Jolla, CA 92093-0319, USA.\\[12pt]
$^{\dagger}$  Theoretical Biology and Biophysics Group, \\
Los Alamos National 
Laboratory, Los Alamos, NM, \break 87545, USA.\\[16pt]
}

\maketitle
\vfill
\eject

\begin{abstract}

The interplay between structure-search of the native structure and
desolvation in protein folding has been explored using a minimalist
model. These results support a folding mechanism where most of the
structural formation of the protein is achieved before water is
expelled from the hydrophobic core. This view integrates water
expulsion effects into the funnel energy landscape theory of protein
folding. Comparisons to experimental results are shown for the SH3
protein. After the folding transition, a near-native intermediate with
partially solvated hydrophobic core is found. This transition is
followed by a final step that cooperatively squeezes out water
molecules from the partially hydrated protein core.
 
\end{abstract}
\eject
\addtocounter{footnote}{1}

The energy landscape theory and the funnel concept combined with a new
generation of experiments provide a new framework for
understanding the protein folding problem. According to this ``new
view,'' the folding landscape for proteins (at least for small fast
folding proteins) resembles a partially rough funnel riddled with
traps where the protein can transiently reside
\cite{landscape,funnel,landscape1,scheraga,dill,eatonrev,gray,shearev}.
Recent theoretical and experimental evidence fully support this
picture and suggests that, since the energetic roughness of the landscape is
sufficiently small (low level of energetic frustration), the
global characteristics of the structural heterogeneity observed in the
transition state ensemble for two-state fast folding proteins are most
strongly determined by the native state topology. For this reason,
theoretical energetically unfrustrated models have been shown to
reproduce nearly all experimental results for the global geometrical
features of the transition state ensemble of a large number of real
proteins, which are two-state or three-state folders, on whose native
structures they are based
\cite{fersht,dobson,oas,roder,englander,baker,serrano,hugh,shoemaker,maritan,gruebele,shea,thirumalai,cecilia,portman,daggett}.

Some concerns have been raised, however, about the applicability of
these models. For example, solvent effects are incorporated implicitly
into the energetic interactions between residues. Such implicit
equilibrium models for solvation are unable to capture effects such as
desolvation of the hydrophobic core; therefore, understanding the
potential of mean force for the interaction between solutes in an
aqueous solvent attracts the attention of many research
groups\cite{chandler1,chandler2,hummer,hummer2,head-gordon,chan}.  The
potential of mean force (PMF) between two methane-like particles
exhibit two minima --- a contact minimum at the van der Waals distance
between the particles and a solvent separated minimum. The two minima
are separated by a desolvation barrier with a width of approximately
one water molecule diameter.  Recently, Hillson {\it et al.}
\cite{garcia2,japan} advertised the importance of the desolvation
effect by taking into account the process of expelling a water
molecule in a pairwise interaction between hydrophobic particles.
Assuming that the time scale for water penetration and escape are
faster than the contact formation in proteins\cite{garcia1}, the shape
of this potential energy function implies that there are free energy
costs for protein de/solvation \cite{hummer}.The time scale separation
assumption may not be always valid and its limitations are discussed
at the following section. The results obtained in these preliminary
studies \cite{garcia2,japan} have been obtained by Monte Carlo
sampling, and therefore they only provide kinetic information
indirectly.

In this article, we generalize these minimalist models to include the
interplay between structure search in a funnel landscape and water
desolvation in determining the protein folding mechanism. Practically,
we provide a phenomenological model where the parameters are adapted
from previous studies by Hummer {\it et al.} \cite{hummer}.  This
features a barrier that plays a dominant role in determining the
desolvation and solvation dynamics during the protein folding
event. Its details are described in the Appendix A.  Applications are
shown for the SH3 protein model, which has a hydrophobic core buried
within beta sheets in the native structure\cite{sh3-rev1}. Our results
show an initial structural collapse to an overall native shape that is
followed by water expulsion from the hydrophobic core region of SH3
during the final step before the native state is reached.  This
folding mechanism is in total agreement with earlier results using energy landscape
theory of folding (e.g, similar ensemble of structures at the
transitions state \cite{cecilia}) with the additional feature that residue-residue
interactions are now generalized to include the possibility of water
solvation/desolvation.

\section{A minimalist model for water mediated interaction}

The interaction potential used is an extension of the energetically
unfrustrated models described above to include the possibility of
protein de/solvation.  This idea is implemented by generalizing $\rm
G\bar{o}$ \cite{go} type interactions; that is, only attractive
interactions are assigned to the native contacts and repulsive
interactions are assigned to the non-native ones.  The
non-native interactions, $U_{rep}(r)$, are given by hard-core
repulsions that follow $ C \times r^{-12}$, where the constant C is optimized
to guarantee a minimally frustrated energy landscape. Again, although
no non-native attractive interactions are included in these
unfrustrated models, what appears to be unrealistic, recent
theoretical and experimental results support their applicability. In
addition to the experimental citations earlier in the introduction,
several recent theoretical/simulational approaches have shown the
geometrical features of the folding mechanism are robust to energetic
frustration as long as it is small relative to the bias towards the
native configuration \cite{Nymeyer2000,Plotkin2000}

 Therefore, even when neglecting the effects of energetic frustration,
these $\rm G\bar{o}$ models have been shown to reproduce nearly all-global
geometrical features of the transition state ensemble and
intermediates of the real, fast folding, proteins
\cite{hugh,shea,cecilia}. The new interactions, however, include not
only the standard minimum for the direct contacts between residues but
also a barrier separating this minimum from a weak secondary one for
the water-separated contacts. As demonstrated in this paper, this
barrier plays a major role in the water solvation/desolvation
mechanism during the kinetic events in the funnel landscape. This
potential is described in details in Fig.~1 and Appendix A.  Three
different regions characterize the nature of contact formation: (a)
when residues (modeled by atoms at the C$_\alpha$ positions) are
separated by a distance shorter than the position of the desolvation
barrier, then a native contact is formed; (b) when this separation is
around the second minimum, i.e., a single water molecule exists
between the residues; defined as a ``pseudo contact'' 
(i.e. the first hydration shell); (c) for larger separations, which
are associated to larger amounts of water between residues, neither
sort of contact is formed.

The two-dimensional free energy profiles plotted as a function of the two 
folding parameters Q and pseudo Q, at the folding temperature ($T_f$, $\frac{k_B 
T_f}{\epsilon}=1.28$), is shown in Fig.~2A and at $0.88 T_f$ in Fig.~2B.  $Q$ 
measures the normalized density of formed native contacts (varying 
between 0 and 1). Pseudo Q is the equivalent parameter for the single water 
separated contacts. Because the desolvation barrier for the interaction 
potential is relatively high (about $1.22 k_B T_f$), it causes some difficulties 
for canonical sampling schemes. This problem can be overcome by incorporating 
multicanonical sampling methods into molecular dynamics simulations 
\cite{method}.  The algorithm for multicanonical sampling 
\cite{multicanonical1,multicanonical4,multicanonical2,multicanonical3}   is 
briefly described in Appendix B.
Figs.~2A and 2B show that the folding mechanism for SH3 is 
governed by two distinct events, although the energy landscape is slightly 
different for the two temperatures. First a folding transition occurs between 
regions around $Q \sim 0.3$ and $Q \sim 0.8$. This step involves 
overcoming a transition state ensemble (TSE) barrier ($Q$ between 0.4 and 0.6). 
At $Q\approx 0.8$, an ensemble of structures with an overall shape similar to 
the folded motif but with substantial internal solvation is reached. This is 
followed by a second transition that is associated to expelling water from the 
hydrophobic core. These hydrated hydrophobic cores, of conformations near the 
native state, have also been observed in free-energy profiles obtained from all-
atom simulations \cite{brooks} where the authors propose that these waters 
provide a lubrication mechanism during folding. This agreement with all-atom 
simulations provides additional support for our simplified water representation. 
These solvated cores are also consistent with experimental evidence discussed 
later.

\section{Kinetic and thermodynamic views of folding mechanism}

In addition to computing the free energy profile, several folding
(unfolding) simulations were performed utilizing Langevin dynamics
simulations \cite{method}. One of the main advantages of these
minimalist representations for proteins is that not only thermodynamic
studies but also a fully kinetic analysis is computationally
feasible. These kinetic runs allow us to fully confirm that the
transition regions identified in the free-energy profile correspond to
the kinetic steps observed during the folding event. Therefore, the
kinetic desolvation events observed during these kinetic events are
exactly the same ones shown in Fig.~2B. To facilitate visualization of
the folding mechanism, a comparison is shown for a typical folding run
at $0.88 \ T_f$ in Figs.~2B, ~2C, ~2D, and~2E. The two transitions
described above can be clearly identified in the kinetic run: the
folding transition in this trajectory occurs after a time $\tau_1$ and
the water expulsion from the hydrophobic core occurs a time $\tau_2$
later. The relative duration of these times may change for different
proteins since both topology and energetics play a major role in
determining the transition barriers. Also, the ratio of $\tau_1$ and
$\tau_2$ depends on the parameterization of the potential of mean
force which may vary with the experimental conditions, such as
pressure \cite{hummer}. In this paper, however, we only consider
systems under near-native conditions. 

To better illustrate these two
transitions, several snapshots along the trajectory are shown in
Fig.~2E.  Unfolded parts of the chain are colored in gray while folded
regions of the chain are shown in red. Pseudo contacts (single water
separated contact pairs) are represented by blue spheres between the
corresponding residues. Unfolded regions with no blue sphere in
between have a larger solvation. The first transition involves a
structural collapse, where the contact formation between the diverging
turn (Dv) and distal loop (Dt) is crucial for assuring the proper
alignment of the tertiary contacts needed to later pack the
hydrophobic core.  These regions overlap with the experimentally
determined high $\phi$-value regions \cite{fersht,sh3-exp1,sh3-exp2}
where contacts are basically formed in the transition states. The next
transition is mainly responsible for the water expulsion out of
hydrophobic core region.  Interestingly, there are 20 pairs of
solvated pseudo contacts distributed about the hydrophobic core and
termini regions.  For visualization purposes, water molecules are
presented as spheres, and the principle of excluded volume is used to
determine whether different pseudo contacts could possibly utilize the
same water molecule. However, we emphasize that no explicit water
molecule are used in the simulation.  After excluding the possibility
that some of these pseudo contacts utilize the same water molecule, we
determined that approximately 17 ``water'' molecules are expelled
cooperatively during this transition. To emphasize this collective
behavior of water expulsion during the interval of the final
transition, an inset in Fig.~2C shows that the number of native pseudo
contacts decreases abruptly, and this decrease matches the number of
contacts being formed.  Most trajectories share similar
characteristics during the folding collapse which is controlled by
topological constraints (the need for the early formation of the
region between Dt and Dv, otherwise folding is not possible) as well
as during the water expulsion mechanism from the hydrophobic core.
This final kinetic feature was absent in the previous minimalist
models using implicit-solvent.  More specifically, quantitative
aspects of this feature are very sensitive on the profile of
desolvation barrier (data not shown).  These results allow us now to
appreciate the relative roles of structure-search towards the native
structure and water desolvation in determining the protein folding
mechanism.

The validity of using the sum of two-body contributions to represent
 the many-body nature of hydrophobic effect is often questioned.
 Czaplewski {\it et al.} \cite{czaplewski} found that three-body
 cooperative term accounts for $10\%$ of the total hydrophobic
 association free energy; it is not too small to be negligible. 
 Shimizu {\it et al.} \cite{shimizu} claimed that, in addition to the
 radial dependence of the potential of mean force, the angular
 dependence of the potential of mean force can modulate the sign of
 the cooperativity.  These results are consistent with earlier observations
by Takada {\it et al.} \cite{takada}. Despite that we can not quantitatively address
 these questions using simplified models, we observe a stronger
 cooperativity in the folding transition utilizing our desolvation
 potential when compared to results utilizing Lennard-Jones
 interactions. This is clear from thermodynamics properties such as
 the specific heat (Fig.~3). Since the enthalphy change for both
 models at individual folding temperatures is basically the same, the
 sharper profile of the specific heat in the desolvation model
 indicates a much stronger cooperativity of folding.  Although an
 increase in the cooperativity of folding is obtained by using a two-body 
 interaction featuring desolvation, such an enhancement is still
 weak to account for the entire cooperativity observed experimentally.

In addition to a sharper peak at the folding temperature in the
specific heat curve obtained for the desolvation model, there is
another distinct thermodynamic signature found at lower temperatures
where a small peak is observed (Fig. 3). This low temperature feature
corresponds to minute fluctuations of the native structure (where
Q$\le 0.93$, Pseudo Q $\le 0.07$).  The origin of these fluctuations
is attributed to few sporadically formed pseudo contacts found at the
termini regions, hydrophobic core, several intra-hairpins, or loops
that are in the proximity of the termini.

\section{Conclusions}

In this study, we have found a near-native intermediate with a partially
solvated hydrophobic core. This intermediate corresponds to a loosely
compact and structured, partially-denatured ensemble recently
identified in experiments for N-terminal SH3 domain of drk using a
N-15/H-2-labeled samples to observe long range amide NOEs
\cite{kay,kay2}. The residues observed experimentally to be partially
hydrated in the vicinity of the core are in good agreement with our
theoretical predictions \cite{kay}. Experimentally, this
``structured'' ensemble was found to have a near-native ``collapsed''
structure that is strongly confined in the conformational space and
close to the native structure; after this state is achieved, folding
required only a simple step to match the well-aligned contact
pairs. In this article we have identified this additional step as a
desolvation process that squeezes out water molecules in the vicinity
of partially hydrated core. More precisely, we have determined that
the folding process of SH3 can be characterized as a
``structure-search collapse'' that is followed by a separate
``desolvation'' step of this nearly-native ensemble.

Is there a biological advantage of having this nearly-native ensemble?
It has been found that this ensemble exists under physiological
conditions and is equally populated compared with native state
\cite{kay2}. Also, experimental evidence suggests that conformational
changes in the core region take place during ligand binding
\cite{bousquet}. We would like to point out that the desolvation
effect might provide a more ``efficient'' (kinetically) and
``inexpensive'' (thermodynamically) way to conduct conformational
changes (i.e., without significant structural rearrangement) and
switch the protein between functional (liganded) and non- functional
states. Interestingly, the ligand binding site of SH3 is far from its
high $\phi$-value regions (such as RT-loop and Dv-turn); thus, the
functional dynamics of SH3 cannot be simply addressed by $\phi$-value
analysis.

In summary, by introducing the feature of desolvation in the tertiary
contacts pair potential, we have been able to capture functional
aspects of SH3 dynamics and folding using a minimalist model.  In
particular, we have identified the nearly-native ensemble that might
have profound biological significance under physiological conditions.
Such a state contributes to the ensemble of conformations existent in
the bottom of the folding funnel.  Although local conformational
changes (specifically de/solvation) of the hydrophobic core is not be
the only mechanism for SH3 liganding and functioning, the combination
of theoretical and experimental evidence strongly suggests its
importance during folding and function of this protein. Despite our 
theoretical studies have been limited to only one
protein, this interplay between structure-search and desolvation
mechanisms is probably general during the folding of different
proteins. The detailed effect on the overall folding mechanism should
vary for different systems, since it will depend on their topological
and/or energetic features.

\section{Appendix}

{\bf A.  Design of the desolvation model}

The desolvation potential, U(r), is designed as a function of the
interacting distance, r, between residues involved in $\rm G\bar{o}$ contacts \cite{go},
where $r'$ is the minimum of the first potential well, $r^{\dag}$ is the
maximum of the desolvation barrier, and $r''$ is the minimum of the
second potential well, respectively.  The separation between $r'$ and
$r''$ is the size of a single water and set to be $3.0
\AA$\cite{garcia2} (see Fig.~1). We set $r^{\dag}=\frac{r'+r''}{2}$.

We required this function to be dependent on several known parameters,
such as the depth of the first well ($\epsilon$), the height of the
desolvation barrier ($\epsilon ''$), the depth of the second well ($\epsilon '$), the
location of two minima and desolvation barrier.  Moreover, we demand
adjustable exponents to control the short and long range
behaviors. The function is therefore segmented into three parts for
better controlling these parameters.

In the equation below, for $r<r'$, we demanded a Lennard-Jones
type of interaction, where $k$ controlled the width of the first
potential well. For $r>r^{\dag}$, we set one boundary to be $r^{\dag}$ such that
$U(r^{\dag})$ has a positive value. Once $r$ increases, $U(r)$ decreases
toward the most negative value at $r=r''$.  Finally, the long tail
behavior is controlled by the power of
$\frac{r^2}{r^{2m}}$. Therefore, $U(r)$ converges to zero when r is
large and $m>2$.  As to $r' < r < r^{\dag}$, we demanded at the junctions
both force and potential have to be continuous.  The function is
therefore segmented into three sections separated by $r'$ and $r^{\dag}$.

\begin{eqnarray}    
{\rm  U(r)=
\begin{cases}   
\epsilon Z(r)(Z(r)-2) \hspace{.5cm} \text{with}  Z(r)=(\frac{r'}{r})^k  \hspace{.5cm} 
\text{if $r<r'$ } \label{eqn.1}\\
C Y(r)^n \frac{\frac{Y(r)^n}{2}-(r^{\dag}-r')^{2n}}{2n} + \epsilon ''
\hspace{.5cm} \text{with} \hspace{.5cm}
\begin{cases}
Y(r)=(r-r^{\dag})^2\nonumber\\
C=\frac{4n(\epsilon+\epsilon '')}{(r^{\dag}-r')^{2n}}\\
\end{cases}  
\text{if $r'\le r<r^{\dag}$ } \\
-B\frac{Y(r)-h_1}{Y(r)^m+h_2} 
\hspace{.5cm} \text{with} \hspace{.5cm}
\begin{cases} 
 Y(r)=(r-r^{\dag})^2\\
 B=\epsilon' m(r''-r^{\dag})^{2(m-1)}\\
h_1=\frac{(1-\frac{1}{m})(r''-r^{\dag})^2}{\frac{\epsilon'}{\epsilon ''}+1}\\ 
h_2=\frac{(m-1)(r''-r^{\dag})^{2m}}{1+\frac{\epsilon ''}{\epsilon'}}\\
\end{cases}
\text{if $r^{\dag} \le r$ } 
\end{cases}
}\end{eqnarray}

{\bf B. Multicanonical sampling method}

In the canonical ensemble at a temperature $T$, the probability
distribution, $P_c$, of the potential energy $E$ is given by
$P_c(E;T) = \frac{1}{Z_c}n(E)e^{-E/k_B T} ,$
with $n(E)$ as the spectral density and $Z_c = \sum_E n(E)e^{-E/k_B T}$.

In our study, the roughness of the energy landscape increases due to
the presence of the desolvation barrier.  Consequently, trapping in
local minima of the energy landscape is easily encountered.  Once sampling 
difficulty emerges, we overcome it by a multicanonical
ensemble sampling techniques. This ensemble is defined by a flat energy 
distribution function, $P_{mc}$:
$P_{mc}(E) = \frac{1}{Z_{mc}}n(E)e^{-W(E)} = constant ,$
where $W(E)$, the weight function, is a function of $E$, and $Z_{mc} = 
\sum_En(E)e^{-W(E)}$.

After the simulations, the reweighting formula converts $P_{mc}$ back
into the canonical distribution, $P_c$, at a temperature $T$
\cite{reweight}:
$P_c(E;T) = \frac{Z_{mc}}{Z_c}P_{mc}(E) e ^{W(E)-E/k_B T} .$
$W(E)$ is not known {\it a priori}, and needs to be refined
through iterated multicanonical simulations at a high temperature
$T_o$ until it covers sufficient wide range of $E$.

\section*{Acknowledgment} 
The authors thank Hugh Nymeyer for assistance in implementing the
multicanonical simulations, and Chinlin Guo for suggestions in the
mathematical modeling of the potential energy function. We also thank
Herbert Levine, Patricia Jennings, John Finke, and Chinlin Guo for
helpful discussions.  This work is supported by the NSF (Grant \#
MCB-0084797). Work at Los Alamos was supported by the LANL Direct
Research Funds (Richard Keller, PI).

\eject

\section*{References and Notes}

\eject

\paragraph*{Fig.~1.}  A schematic representation of a phenomenological
potential for tertiary contact formation which includes the
possibility of desolvation.  $r'$ and $r''$ label the residue-residue
contact minimum and the single water molecule-separated contact
minimum, respectively.  The parameters for energies are adapted from
the references\cite{hummer,garcia2}, where
$\frac{\epsilon^{``}-\epsilon^{'}}{\epsilon^{'}-\epsilon}=1.33$,
$\frac{\epsilon^{'}}{\epsilon}=\frac{1}{3}$.  The desolvation barrier
allows for a clear determination of whether a native contact is formed
or not.  Three regions are defined: (a) when the separation distance
between residues $r$ is shorter than the range of the desolvation
barrier, then a native contact is formed; (b) when the residues are
separated from the desolvation barrier by a single water
molecule($3\AA$) distance, then a ``pseudo contact'' (i.e. the first
hydration shell) is formed; (c) when the residues are separated by
multiple hydration shells, no contact is formed.  By using this simple
rule, the degree of nativeness of any configuration is easily
determined.  The normalized density of the native contacts ($Q$) and
the native ``pseudo contacts'' (pseudo $Q$) are appropriate parameters
to characterize the degree of nativeness and of desolvation.  For
comparison, a Lennard-Jones type potential (in red) is presented in
the figure.

\paragraph*{Fig.~2.}

Panel A and B show the free energy diagrams plotted as the function of
folding parameters $Q$ and pseudo $Q$ at the folding temperature
($T_f$, where $\frac{k_B T_f}{\epsilon}=1.28$) and 0.88 $T_f$,
respectively.  $Q$ and pseudo $Q$ are the normalized density of native
contacts and pseudo native contacts (see Fig. 1).  Free energies are
measured in units of $k_B T$ ($T$=$T_f$ in panel A or 0.88$T_f$ in
panel B) where blue(red) stands for low(high) value of them.  Pseudo
$Q$ ranges approximately between $0$ and $0.3$.  Pseudo $Q$ appears to
decline sharply at $Q\ge 0.8$, indicating the system expels water at
this stage.  A typical folding trajectory is superimposed on the free
energy landscape at $T=0.88T_f$ (panel D).  In addition, $Q$ and
pseudo $Q$ (panel C), and contact energy (panel E) are plotted as a
function of integration time step.  Using the same desolvation
potential presented in Fig.~1, this trajectory is simulated by
Langevin dynamics.  The folding run was recorded for another 10
millions steps after $Q$ reaches 1.  According to the distinct
transitions indicated by the contact energy in (panel E), the folding
mechanism is characterized by two stages: a structural collapse
towards a nearly-native ensemble at a time $\tau_1$ which is followed
by water expulsion from the hydrophobic core at a time $\tau_2$.  To
have a better understanding of the kinetics, several snapshots of the
chain are shown in which a blue sphere is used to identify pseudo
contacts (i.e. single water molecule-separated residues).  In
addition, we color the residues with formed native contacts in red to
specify the folded portion of the protein and the unfolded portion in
gray.  In this trajectory, (a) shows an unfolded configuration where
only the short-range native contacts are formed.  This early
step is followed by a transition indicating the structural collapse to
the nearly-native ensemble (b $\to$ c) in which $Q$ increases and pseudo 
$Q$ decreases. The configuration
shown in (b) has 31 pseudo contacts formed.  After excluding the
possibility that some of these pseudo contacts utilize the same water
molecules, 23 ``water" molecules are expelled cooperatively during
this transition (the principle of excluded volume is used to determine
whether pseudo contacts possibly utilize the same water molecule).
Notably, the native contact formation between the diverging turn (Dv)
and the distal loop (Dt) (indicated by a broken green circle) is
crucial for this structural collapse.  This region overlaps with the
experimentally determined high $\phi$-value
regions\cite{sh3-exp1,sh3-exp2}.  The final transition involves the
mechanism of water expulsion from the partially hydrated hydrophobic
core (d $\to$ e). This transition involves the formation
of long-range tertiary contacts across the two sandwiched
$\beta$-sheets.  Configuration (d) has 20 pseudo contacts in the
hydrophobic core and the terminal regions.  After excluding the
possibility that some of these pseudo contacts utilize the same water
molecules, 17 ``water" molecules are expelled cooperatively during
this final transition.  Configuration (e) has only a few residual
``water" molecules in the terminal regions.  To emphasize this
collective behavior of water expulsion during the interval of the
final transition, an inset in panel C shows that the number of native
pseudo contacts decreases abruptly (N is the number of total native
contacts).

\paragraph*{Fig.~3.}

The cooperativity of folding of models using the desolvation and
Lennard-Jones potentials are compared by plotting specific heat as a
function of temperature.  The enthalphy change for both models at
individual folding temperatures is the same. The sharper profile of
the specific heat implies stronger cooperativity of folding in the
desolvation model than that for the Lennard-Jones model.  Noticeably,
at low temperature, there is a broad but small peak in the desolvation
model.  This low temperature feature corresponds to minute
fluctuations of the native structure (where Q$\le 0.93$, Pseudo Q $\le
0.07$).  The origin of the fluctuation at low temperature is
attributed to very few sporadically formed pseudo contacts found at
the termini regions, hydrophobic core, several intra-hairpins, or
loops that are in the proximity of the termini. In the inset with an
SH3 structure, the top $30\%$ residues that possibly form pseudo
contacts at low temperatures are highlighted with purple balls.  In
comparison, the regions colored in red are the binding site of the
protein to the proline-rich peptide.  The green broken circles label
the experimentally determined high $\phi$-value
regions\cite{sh3-exp1,sh3-exp2}. The beta sheet colored in cyan is the
region, in addition to termini and loops regions, suggested by the NMR
experiments\cite{kay,kay2} to account for the conformational changes
under near-native conditions as temperature mildly increases.


\eject

\begin{figure}\label{PMF}
\centerline {\vbox{
{\epsfxsize = 8.7cm \epsffile{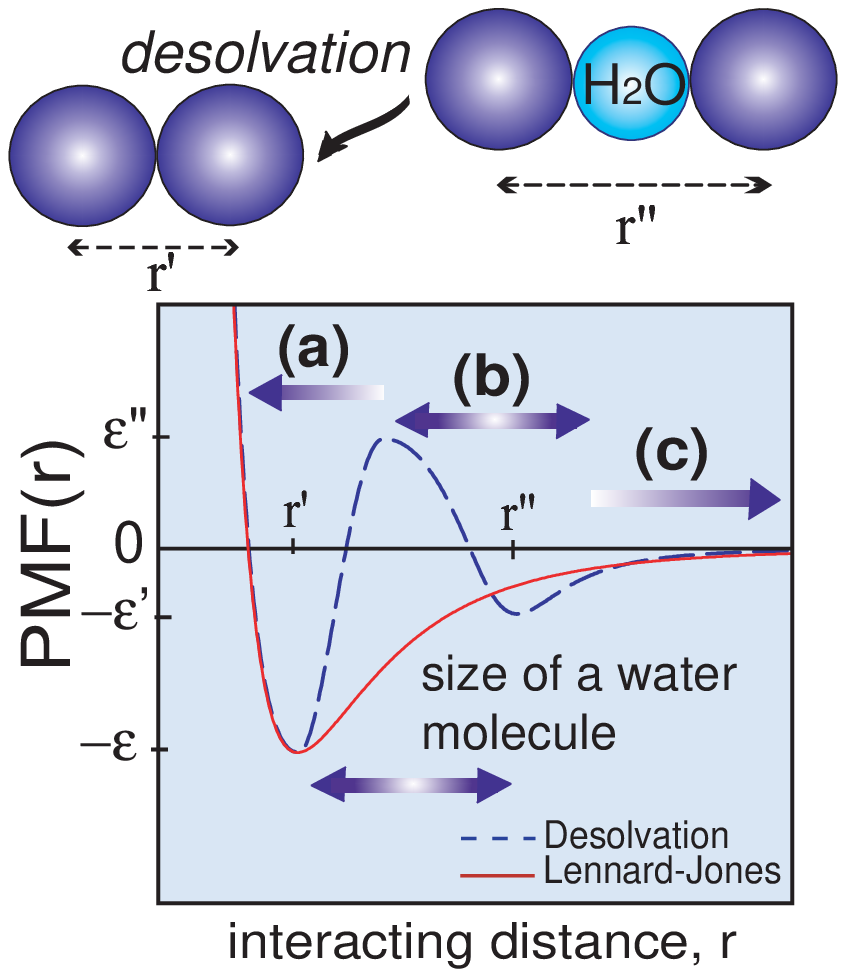}}
}}  
\caption{}
\end{figure}

\begin{figure}\label{2D}
\centerline {\vbox{
{\epsfxsize = 18.0cm \epsffile{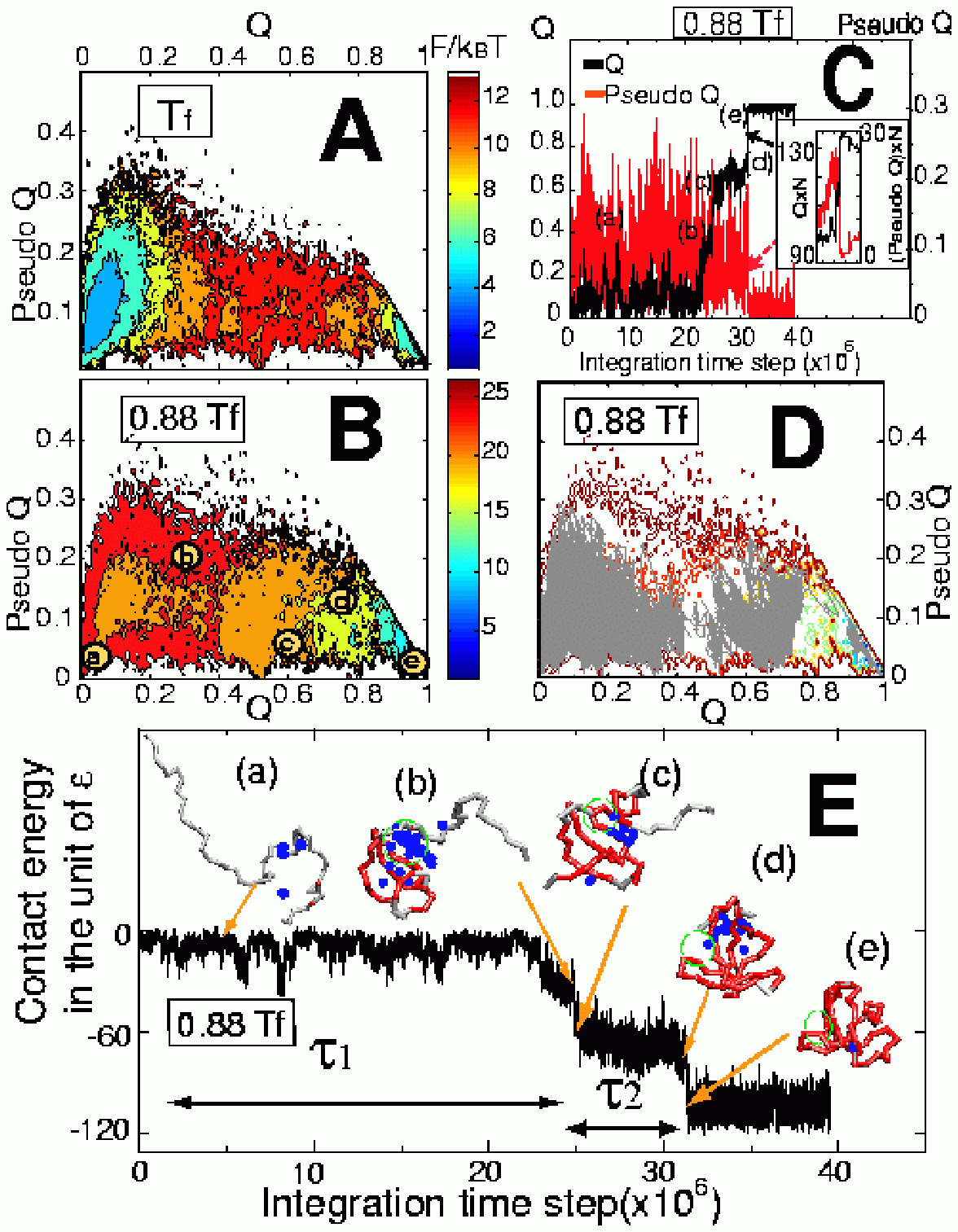}}
}}  
\caption{}
\end{figure}

\begin{figure}\label{Cv}
\centerline {\vbox{
{\epsfxsize = 8.7cm \epsffile{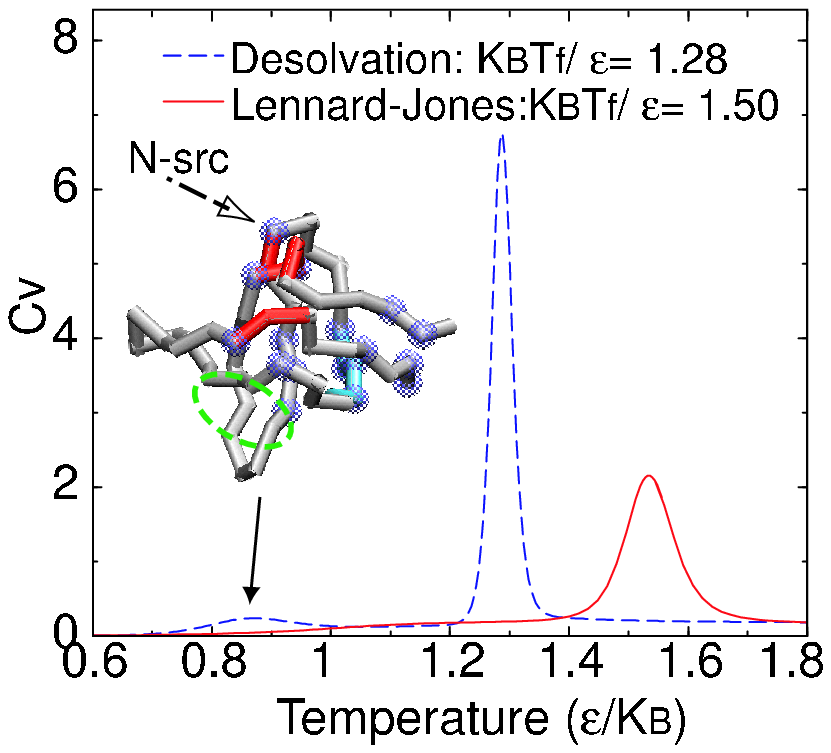}}
}}  
\caption{}
\end{figure}


\begin{enumerate}
\bibitem{landscape}Bryngelson,~J.~D.,~Wolynes,~P.~G.(1987) {\it Proc. Natl. 
Acad. 
Sci.} U.S.A. {\bf 84}, 7524-7528.
\bibitem{funnel}Leopold,~P.~E.,~Montal,~M., \&~Onuchic,~J.~N.(1992)
{\it Proc. Natl. Acad. Sci.} U.S.A. {\bf 89}, 8721-8725.
\bibitem{landscape1}Onuchic,~J.~N., Luthey-Schulten,~Z., 
\&~Wolynes,~P.~G.(1997) 
{\it Annu. Rev. Phys. Chem.} {\bf 48}, 545-600.
\bibitem{scheraga}
Scheraga,~H.~A.(1992)
{\it Protein Sci.} {\bf 1}, 691-693.
\bibitem{dill}
Dill,~K.~A., Chan,~H.~S.(1997)
{\it Nature Struct. Biol.} {\bf 4}, 10-19.
\bibitem{eatonrev}Eaton,~W.~A.,~Mu\~noz,~V.,
~Thompson,~P.~A.,~Chan,~C.-K., \&~Hofrichter,~J.(1997)
{\it Curr. Opin. Struct. Biol.} {\bf 7}, 10-14.
\bibitem{gray}Telford,~J.~R.,~Wittung-Stafshede,~P.,~Gray,~H.~B., 
\&~Winkler,~J.~R.(1998) {\it Acc. Chem. Res.} {\bf 31}, 755-763.
\bibitem{shearev}Shea,~J.-E.,~Brooks,~III,~C.~L.(2001)
 {\it Annu. Rev. Phys. Chem.} {\bf 52}, 499-535.
\bibitem{fersht}
Jackson,~S.~E.,~elMasry,~N., \&~Fersht,~A.~R.(1993) {\it Biochemistry} 
{\bf 32}, 11270-11278.
\bibitem{dobson}
Plaxco,~K.~W.,~Dobson,~C.~M.(1996)
{\it Curr. Opin. Struct. Biol.} {\bf 6}, 630-636.
\bibitem{oas}
Burton,~R.~E.,~Huang,~G.~S.,~Daugherty,~M.~A.,~Calderone,~T.~L., \&~Oas,~T.~G.
(1997) {\it Nature Struct. Biol.} {\bf 4}, 305-310.
\bibitem{roder}
Roder,~H.,~Colon,~W.(1997)
{\it Curr. Opin. Struct. Biol.} {\bf 7}, 15-28.
\bibitem{englander}
Sosnick,~T.~R.,Shtilerman,~M.~D.,~Mayne,~L., \&~Englander,~S.~W.(1997)
{\it Proc. Natl. Acad. Sci.} U.S.A. {\bf 94}, 8545-8550.
\bibitem{baker} 
Riddle,~D.~S.,~Grantcharova,~V.~P.,~Santiago,~J.~V.,~Alm,~E.,~Ruczinski,~I., 
\&~Baker,~D.
(1999) {\it Nature Struct. Biol.} {\bf 6}, 1016-1024.
\bibitem{serrano}
Mart\'{\i}nez,~J.~C.,~Serrano,~L.(1999) {\it Nature Struct. Biol.}
{\bf 6}, 1010-1016.
\bibitem{hugh}
Nymeyer,~H.,~Garc\'{\i}a,~A.~E., \&~Onuchic,~J.~N.(1998)
{\it Proc. Natl. Acad. Soci.} U.S.A. {\bf 95}, 5921-5928.
\bibitem{shoemaker}
Shoemaker,~B.~A.,~Wolynes,~P.~G.(1999)
{\it J. Mol. Biol.} {\bf 287}, 657-674.
\bibitem{maritan}
Micheletti,~C.,~Banavar,~J.,~Maritan,~A., \&~Seno,~F.(1999)
{\it Phys. Rev. Letters} {\bf 82}, 3372-3375.
\bibitem{gruebele}
Gruebele,~M.(1999)
{\it Annu. Rev. Phys. Chem.} {\bf 50}, 485-516.
\bibitem{shea}
Shea,~J.-E.,~Brooks,~III,~C.~L., \&~Onuchic,~J.~N.(1999)
 {\it Proc. Natl. Acad. Sci.} U.S.A. {\bf 96}, 12512-12517.
\bibitem{thirumalai}Klimov,~D.~K.,~Thirumalai,~D.(2000) 
{\it Proc. Natl. Acad. Soci.} U.S.A. {\bf 97}, 7254-7259.
\bibitem{cecilia}
Clementi,~C.,~Nymeyer,~H., \&~Onuchic,~J.~N.(2000) 
{\it J. Mol. Biol.} {\bf 298}, 937-953.
\bibitem{portman}
Portman,~J.~J.,~Takada,~S., \&~Wolynes,~P.~G.(2001)
{\it J. Phys. Chem.} {\bf 114}, 5069-5081.
\bibitem{daggett}
Pan,~Y.~P.,~Daggett,~V.(2001)
{\it Biochemistry} {\bf 40}, 2723-2731.
\bibitem{chandler1}
Pratt,~L.,~Chandler,~D.(1980)
{\it J. Chem. Phys.} {\bf 73}, 3434-3441.
\bibitem{chandler2}
Lum,~K.,~Chandler,~D., \&~Weeks,~J.~D.(1999)
{\it J. Phys. Chem. B} {\bf 103}, 4570-4577.
\bibitem{hummer}
Hummer,~G.,~Garde,~S.,~Garc\'{\i}a,~A.~E.,~Paulaitis,~M.~E., 
\&~Pratt,~L.~R.(1998) {\it 
Proc. Natl. Acad. Sci.} U.S.A. {\bf 95}, 1552-1555.
\bibitem{hummer2}
Hummer,~G.,~Garde,~S.,~Garc\'{\i}a,~A.~E.,~Pohorille,~A., \&~Pratt,~L.~R.(1996) 
{\it  Proc. Natl. Acad. Sci.} U.S.A. {\bf 93}, 8951-8955.
\bibitem{head-gordon}J.~M.~Sorenson,~J.~M.,~Hura,~G.,~Soper,~A.~K.,~Pertsemlidis,~A., \&~Head-
Gordon,~T.(1999)
{\it J. Phys. Chem. B} {\bf 103}, 5413-5426.
\bibitem{chan}
Shimizu,~S.,~Chan,~H.~S.(2000)
{\it J. Phys. Chem.} {\bf 113}, 4683-4700.
\bibitem{garcia2}
Hillson,~N.,~Onuchic,~J.~N., \&~Garc\'{\i}a,~A.~E.(1999) 
{\it Proc. Natl. Acad. Sci.}
U.S.A. {\bf 96}, 14848-14853.
\bibitem{japan}
Garc\'{\i}a,~A.~E.,~Hillson,~N., \&~Onuchic,~J.~N. (2000) {\it
Prog. Theor. Phys. Supp.} {\bf 138}, 282-291.
\bibitem{garcia1}
~Garc\'{\i}a,~A.~E.,~Hummer,~G.(2000) {\it Proteins} {\bf 38}, 261-272.     
\bibitem{Nymeyer2000} Nymeyer,~H.,Socci,~N.~D., \&~Onuchic,~J.~N.(2000) {\it 
Proc. Natl. Acad. Sci.} U.S.A. {\bf 97}, 634-639.
\bibitem{Plotkin2000} Plotkin,~S.~S., \&~Onuchic,~J.~N.(2000) {\it Proc. Natl. 
Acad. Sci.} U.S.A. {\bf 97}, 6509-6514.
\bibitem{sh3-rev1}
Musacchio,~A.,~Noble,~M.,~Pauptit,~R.,~Wierenga,~R., \&~Saraste,~M.(1992) 
{\it Nature} {\bf 359}, 851-855.
\bibitem{go}
Ueda,~Y.,~Taketomi,~H., \&~$\rm G\bar{o}$,~N.(1978) 
{\it Biopolymers} {\bf 17}, 1531-1548.
\bibitem{method} 
(a)Molecular Dynamics (MD): we employ standard molecular dynamics
method carried out by the AMBER6 program as an integrator.  The
dynamics of the system is studied at constant temperature with
the Berendsen algorithm\cite{berendsen}.  To ensure thorough sampling, the
dynamics is simulated for sufficiently long runs (over $1\times 10^8$ integration
time steps) in either canonical or multicanonical ensemble.  In
either ensemble, over $1\times 10^5$ configurations are collected.
(b)Langevin Dynamics (LD): To study kinetics, we simplify the system
as large-sized residues in water such that the limit of high
frictional region is appropriate. 
Several independent folding runs, initiated from various
high-temperature configurations, are performed at the various
temperatures.  
(c)The unit of energy is $\epsilon$.  
The averaged bond length is $\sigma$. 
The monomer mass is m. Time is
measured in the units of reduced time $\tau_s =
\left(\frac{m\sigma^2}{\epsilon}\right)^{\frac{1}{2}}$.  The
integration time is 0.005$\tau_s$.  Temperature is measured in units
of $\left(\frac{\epsilon}{k_B}\right)$, where $k_B$ is the Boltzmann
constant.
\bibitem{multicanonical1}
Berg,~B.~A.,~Neuhaus,~T.(1992)
{\it Phys. Rev. Lett.} {\bf 68}, 9-12.
\bibitem{multicanonical4}
Hansmann,~U.~H.~E.,~Okamoto,~Y., \&~Eisenmenger,~F.
(1996){\it Chem. Phys. Lett} {\bf 259}, 321-330.
\bibitem{multicanonical2}
Nakajima,~N.,~Nakamura,~H., \&~Kidera,~A.(1997) {\it J. Phys. Chem. B}  
 {\bf 101}, 817-824.
\bibitem{multicanonical3}
Mitsutake,~A.,~Sugita,~Y., \&~Okamoto, Y.(2001) {\it Biopolymers}
 {\bf 60}, 96-123.
\bibitem{reweight}
Ferrenberg,~A.~M., \&~Swendsen,~R.~H. (1988)  {\it Phys. Rev. Lett.}
{\bf 61}:2635-2638
\bibitem{brooks}
Sheinerman,~F.~B.,~Brooks, III,~C.~L. (1998)
{\it J. Mol. Biol.}  {\bf 278}, 439-456.
\bibitem{berendsen}
Berendsen,~H.~J.~C.,~Postma,~J.~P.~M.,~van Gunsteren,~W.~F.,~DiNola,~A., 
\&~Haak,~J.~R.(1984)  {\it J. Chem. Phys.} {\bf 81}, 3684-3690.
\bibitem{sh3-exp1} 
~Grantcharova,~V,~P.,~Riddle,~D.~S.,~Santiago,~J.~V., \&~Baker,~D.(1998)
{\it Nature Struct. Biol.} {\bf 5}, 714-720.
\bibitem{sh3-exp2}
Mart\'{\i}nez,~J.~C.,~Pisabarro,~M.~T., \&~Serrano,~L.(1998) {\it Nature Struct. 
Biol.}{\bf 5}, 721-729.
%
\bibitem{czaplewski}
Czaplewski,~C.,~Rodziewicz-
Motowidlo,~S.,~Liwo,~A.,~Ripoll,~D.~R.,~Wawak,~R.~J.,~\&~Scheraga,~H.~A.(2000)
{\it Protein Sci.} {\bf 9},1235-1245.
%
\bibitem{shimizu}
Shimizu,~S.,~Chan,~H.~S.(2001)
{\it J. Chem. Phys.} {\bf 115}, 1414-1421.
\bibitem{takada}
Takada,~S.,~Luthey-Schulten,~Z., \&~Wolynes,~P.~W.(1999)
{\it J. Chem. PHys.} {\bf 110}, 11616-11629.
%
\bibitem{kay}
Zhang,~O.,~Forman-Kay,~J.~D.(1997)
{\it Biochemistry} {\bf 36}, 3959-3970.
\bibitem{kay2}
Mok,~Y.-K.,~Kay,~C.~M.,~Kay,~L.~E., \&~Forman-Kay,~J.(1999)
{\it J. Mol. Biol.} {\bf 289}, 619-638.
\bibitem{bousquet}
Bousquet,~J.~A.,Garbay,~C.,~Roques,~B.~P., \&~M\'ely,~Y.(2000)
{\it Biochemistry} {\bf 39}, 7722-7735.
\end{enumerate}
\end{document}